\documentclass[aps,prl,reprint,superscriptaddress,intlimits]{revtex4-2}
\usepackage{bm,latexsym,mathrsfs,enumerate,amsmath,amssymb}

\usepackage{graphicx}
\usepackage[breaklinks=true,urlcolor = blue,colorlinks = true,citecolor = blue,linkcolor = blue]{hyperref}
\usepackage[utf8]{inputenc}
\usepackage{siunitx}

\begin{document}

\title{Creating and Driving a Twist Soliton on a Magnetic Skyrmion Tube}

\author{Shoya Kasai}
  \email{Contact author: kasai@aion.t.u-tokyo.ac.jp}
  \affiliation{Department of Applied Physics, The University of Tokyo, Hongo, Tokyo 113-8656, Japan}
\author{Kotaro Shimizu}
  \affiliation{Department of Applied Physics, The University of Tokyo, Hongo, Tokyo 113-8656, Japan}
\author{Shun Okumura}
  \affiliation{Quantum-Phase Electronics Center (QPEC), The University of Tokyo, Tokyo 113-8656, Japan}
  \affiliation{RIKEN Center for Emergent Matter Science (CEMS), Wako, Saitama 351-0198, Japan}
  \affiliation{International Institute for Sustainability with Knotted Chiral Meta Matter (WPI-SKCM$^2$), Hiroshima University, Hiroshima 739-8531, Japan}
\author{Yukitoshi Motome}
  \email{Contact author: motome@ap.t.u-tokyo.ac.jp}
  \affiliation{Department of Applied Physics, The University of Tokyo, Hongo, Tokyo 113-8656, Japan}

\begin{abstract}
A magnetic skyrmion tube is a three-dimensional topological soliton formed by stacking two-dimensional skyrmions along the out-of-plane direction. Recent real-space observations of skyrmion tubes have stimulated growing interest in their dynamics and emergent properties. Here, we go beyond simple skyrmion stacking and investigate how a ``twist" introduced along the tube direction affects the dynamics and emergent responses of skyrmion tubes. We find that such a twist can be created as a localized texture, termed a twist soliton, through thermal quench dynamics. By complementarily combining large-scale numerical simulations with analytical calculations based on collective coordinates, we clarify its current-driven nonlinear motions that depend on its twist chirality. Remarkably, its velocity can be substantially enhanced by a magnetic-field component perpendicular to the tube. Furthermore, the associated emergent electric field enables identification of the twist soliton, including the sign of its chirality, through Hall measurements. Our results reveal the twist degree of freedom as an essential ingredient of skyrmion-tube physics and pave the way for the development of spintronic devices exploiting the three-dimensional nature of spin textures.
\end{abstract}
\maketitle
% \tableofcontents

%%%%%%%%%%%%%%%%%%%%%%%%%%%%%%%%%%%%%%%%%%%%%%%%%%%%%%%%%%%%%%%%%%%%%%%%%%%%%%%%%%%%%%%%%%%%%%%%%%%%%%%%%%%%%%%%%%%%%%%%%%%%%%%%%%%%

Magnetic media can host topological solitons unique to one-, two-, and three-dimensional geometries, whose stability, dynamics, and emergent electromagnetic responses have attracted broad interest. For example, the current-driven dynamics of one-dimensional (1D) domain walls and two-dimensional (2D) skyrmions have inspired a variety of potential spintronic applications~\cite{Freitas1985,Slonczewski1996,Berger1996,Yamaguchi2004,Klaui2005,Barnes2006,Barnes2007,Duine2008,Tserkovnyak2008,Parkin2008,Yang2009,Shibata2011,Jonietz2010,Jiadong2011,Schulz2012,Fert2013,Iwasaki2013_comm,Iwasaki2013_tech,Nagaosa2013,Koshibae2015,Jiang2017}. With recent progress in three-dimensional (3D) magnetization imaging~\cite{Park2014,Donnelly2017,Seki2022,Wolf2022,Yu2022,Henderson2023,Zheng2023,Yu2024}, research on topological magnetism is now extending from well-established 1D and 2D textures to fully 3D systems.

The simplest solitonic texture in 3D magnets is the skyrmion tube, formed by stacking skyrmions along the out-of-plane direction~\cite{Milde2013,Birch2020}. Other examples such as chiral bobbers~\cite{Zheng2018}, torons~\cite{Leonov2018}, and hopfions~\cite{Sutcliffe2018} can be viewed as its variants. Owing to the extension of the skyrmion into the third spatial dimension, skyrmion tubes possess additional deformation modes, such as bending and twisting. Various approaches to induce the tube bending have been investigated, unveiling the nonlinear Hall responses and collective spin excitations~\cite{Yokouchi2018,Shizeng2019,Seki2020,Xiangjun2020,Okumura2023,Kravchuk2023}. Such flexibility also allows solitary wave propagation along a tube, analogous to the Hasimoto soliton on a fluid vortex filament~\cite{Hasimoto1972,Kravchuk2020}. In contrast, the role of tube twisting in dynamics and response phenomena remains largely unexplored.

\begin{figure}[b!]
  \centering
  \includegraphics[width=\hsize]{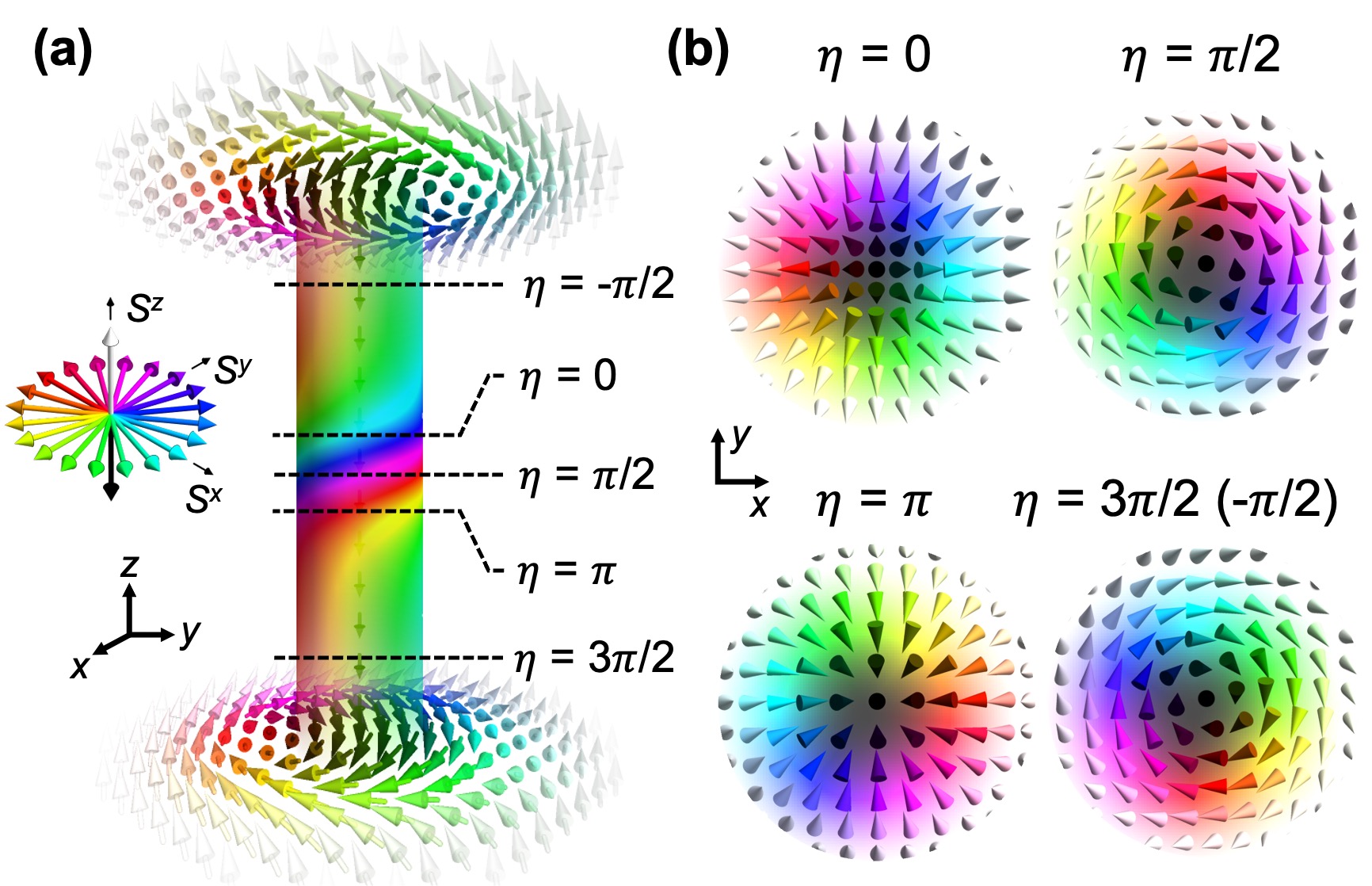}
  \caption{(a) Schematic illustration of a locally twisted magnetic skyrmion tube with twist chirality $\chi = -1$. Cross sections parallel to the $xy$ plane host a magnetic skyrmion with $z$-dependent helicity $\eta$, which varies continuously by $2\pi \chi$ from bottom to top. The color code shown in the left inset is used throughout this paper to represent spin configurations. (b) Skyrmions at the cross sections represented by the dashed lines in (a) with helicities $\eta = 0, \pi/2, \pi$, and $3\pi/2\,(-\pi/2)$.}
  \label{schematic}
\end{figure}

Here, the twist refers to a modulation of the skyrmion helicity $\eta$ along the tube direction, as illustrated in Fig.~\ref{schematic}(a). The helicity determines the swirling pattern of a skyrmion, as shown in Fig.~\ref{schematic}(b), and is typically fixed by spin interactions such as Dzyaloshinskii-Moriya interactions (DMI) and dipole-dipole interaction (DDI)~\cite{Tokura2021}. Thus, skyrmion tubes are often assumed to have uniform helicity, and intrinsic twisting has received limited attention, except for an extrinsic mechanism induced by interfacial effects~\cite{Meynell2014,Kawaguchi2016,Leonov2016,Brearton2022,Kong2024}. Recent theoretical studies, however, have shown that twisted skyrmion tubes can be stable even without surface effects~\cite{Yokota2021, Saji2025}. In particular, a localized $\pi$ twist was found to be stabilized by the DDI, reflecting the twofold helicity degeneracy in such systems~\cite{Yokota2021}. This can be viewed as a topological soliton intrinsic to the twist degree of freedom. Since the DMI instead selects a single helicity, it may support a distinct type of twisted structure, but this possibility remains unexplored. Given the rich phenomena uncovered for bending skyrmion tubes, exploring twisted textures and their dynamics may reveal further intriguing responses in 3D topological magnetism.

In this Letter, we numerically investigate the creation and current-driven dynamics of a $2\pi$-twisted skyrmion tube in a 3D chiral magnet. We show that a localized $2\pi$ twist, termed a ``twist soliton", is probabilistically generated through a thermal quench. We further demonstrate that an in-plane electric current induces not only linear transverse motion within the plane but also nonlinear out-of-plane motion of the twist soliton, whose velocity can be widely tuned by an in-plane magnetic field. Finally, we propose a detection scheme for the twist solitons based on Hall voltage measurements of their associated emergent electric field. Our findings provide a foundation for exploring twist solitons intrinsic to the 3D nature of skyrmion tubes.

We consider a 3D chiral magnet, incorporating competing exchange interactions, the DMI, and a Zeeman coupling. Spin dynamics is investigated by solving the Landau-Lifshitz-Gilbert (LLG) equation in Eq.~\eqref{eq:LLG} including spin-transfer torque (STT) in Eq.~\eqref{eq:STT}. Details of the model and numerics are provided in the End Matter.

\begin{figure}[t!]
  \centering
  \includegraphics[width=\hsize]{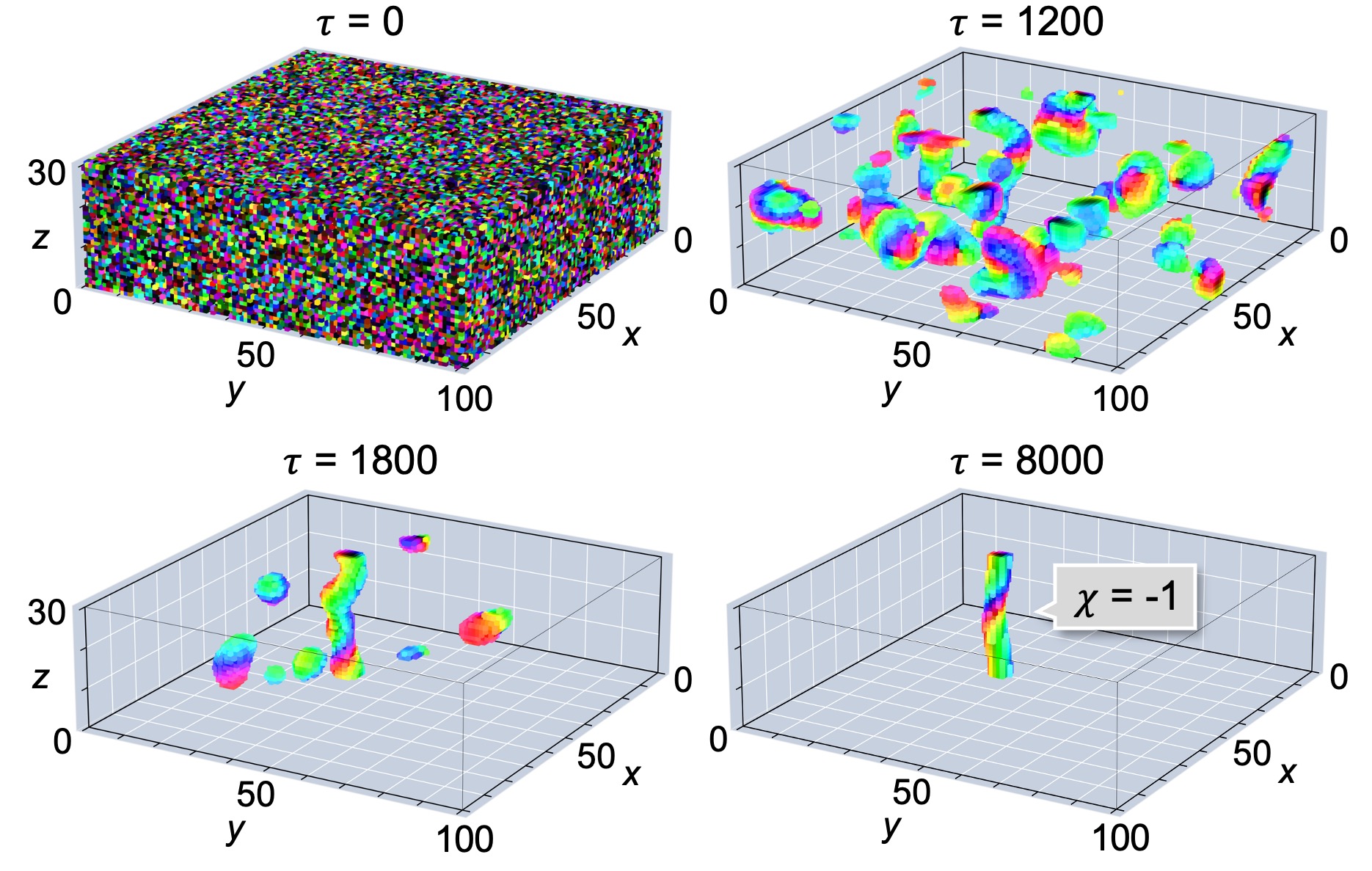}
  \caption{Formation process of a twisted skyrmion tube with chirality $\chi=-1$ for $(B_z, K) = (0.01, 0.01)$. Only spins satisfying $S^z \lesssim 0$ are shown. The system size is $N = 101^2 \times 31$ under periodic boundary conditions. Each panel presents the time evolution of the spin configuration obtained from a zero-temperature LLG simulation starting from a random spin state.}
  \label{quench}
\end{figure}

First, motivated by the Kibble-Zurek mechanism (KZM)~\cite{Kibble1976, Zurek1985}, a general theory of defect formation during thermal quench, we demonstrate that skyrmion tubes with a $2\pi$ twist can be created during relaxation dynamics starting from a random spin state. It has been suggested that the KZM is applicable for magnetism, exemplified by formation of skyrmions~\cite{Shizeng2016} and skyrmion tubes without twists in a spin model different from ours~\cite{Kuchkin2025}. To enhance the probability of generating skyrmion tubes, we consider a thin-plate geometry with a system size $N = 101^2 \times 31$ and add a weak easy-axis anisotropy $K = 0.01$. Figure~\ref{quench} shows snapshots of the relaxation dynamics without an electric current $j_{\rm e} = 0$ in Eq.~\eqref{eq:LLG}, starting from a random spin state corresponding to the high-temperature limit at time $\tau = 0$. The subsequent relaxation is performed at zero temperature.

As shown in the snapshots from $\tau = 1200$ to $1800$, the magnetic field gradually aligns spins upward, eventually leaving a single tubelike texture at $\tau = 8000$. This remnant is identified as a $2\pi$-twisted skyrmion tube with twist chirality $\chi = -1$, as schematically shown in Fig.~\ref{schematic}(a). The twisted region is spatially localized, which we attribute to the present choice of the DM vector: it favors the helicity $\eta = 3\pi/2$ and thus suppresses the broad distribution of the twist. This behavior resembles helimagnetism under an external magnetic field, where a spin helix evolves into a chiral soliton~\cite{Togawa2012}. We therefore refer to the localized twist on a tube as a ``twist soliton". We performed 500 independent simulations and confirmed that several types of twisted tubes are generated probabilistically during the relaxation process (see End Matter for details).

\begin{figure}[t!]
  \centering
  \includegraphics[width=\hsize]{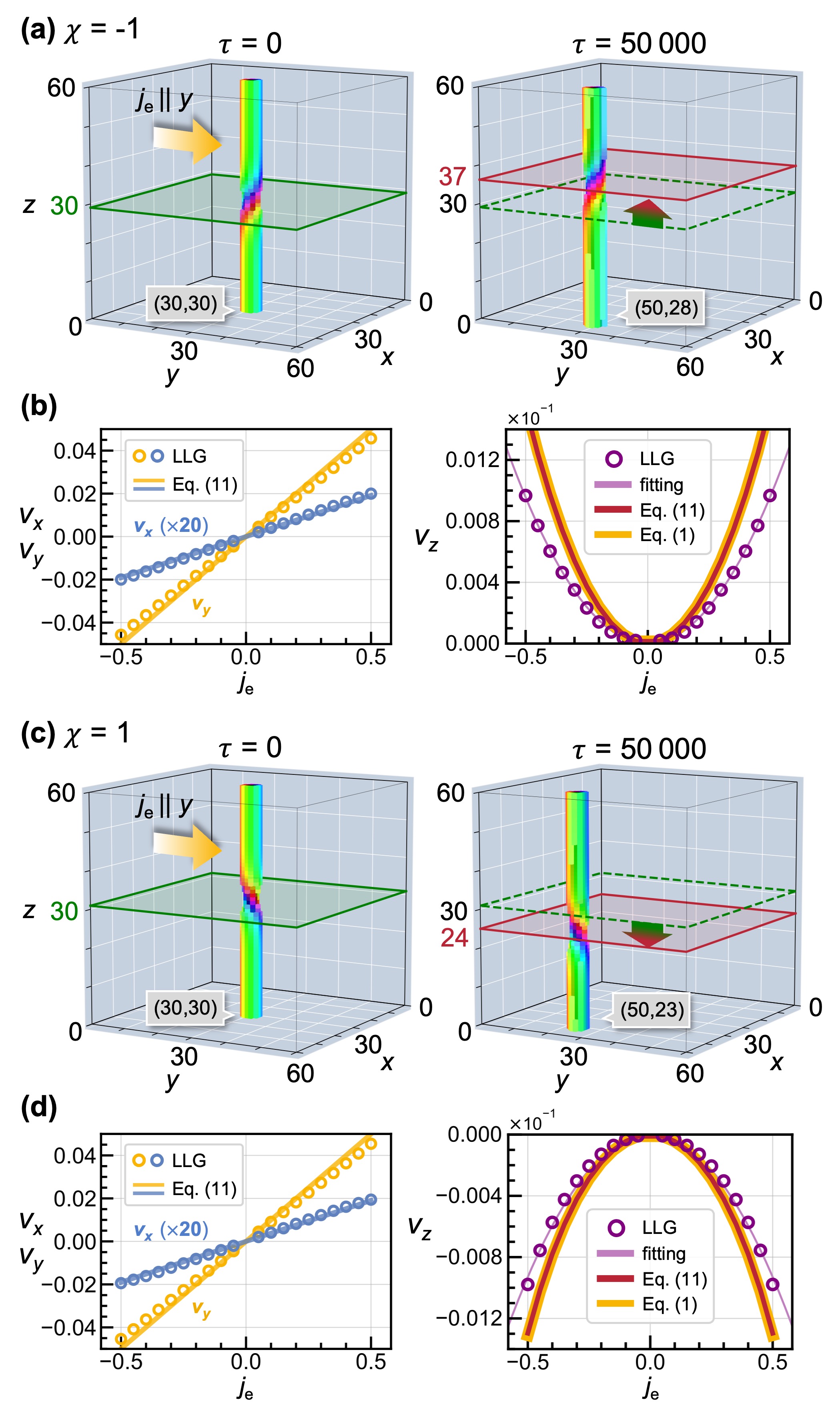}
  \caption{In-plane current-induced dynamics of twisted skyrmion tubes and its dependence on $j_{\rm e}$ for (a,b) $\chi= -1$ and (c,d) $\chi= 1$. The left and right panels of (a,c) show the initial and final states with $(B_z, j_{\rm e}) = (0.02, 0.2)$, respectively. The green and red planes mark the twist positions in the initial and final states, respectively. The left panels of (b,d) show the twisted tube velocity $v_x$ and $v_y$ as functions of $j_{\rm e}$. The values of $v_x$ are multiplied by $20$ for visibility. The circles denote simulation results, and the solid lines are predictions from Eq.~\eqref{eq:thiele}. The right panels of (b,d) show the twist soliton velocity $v_z$. The purple solid lines are quadratic fits $\propto j_{\rm e}^2$ to the numerical results plotted by the purple circles. The red and orange solid lines indicate the predictions from Eqs.~\eqref{eq:thiele} and \eqref{eq:nonlinearvz}, respectively.}
  \label{DCjey}
\end{figure}

Next, to clarify the effect of the twist soliton and its chirality on the skyrmion tube dynamics, we investigate 3D motion driven by an electric current perpendicular to the tube with $\hat{\bm{\nu}} = \hat{\bold{y}}$ in Eq.~\eqref{eq:STT}. We hereafter consider an isotropic system with $N = 61^3$ and $K = 0$. We have confirmed that a nonzero $K$ does not qualitatively change the dynamics. Figures~\ref{DCjey}(a) and \ref{DCjey}(c) show snapshots of twisted tubes with $\chi = -1$ and $\chi = 1$, respectively, for $(B_z,j_{\rm e}) = (0.02, 0.2)$. The initial states shown on the left are prepared by performing energy minimization on a skyrmion ansatz~\cite{Nagaosa2013}; see Ref.~\cite{Kasai2024} for numerical details. The tube centers are initially placed at $(x,y)=(30,30)$, and the green planes mark the initial position of the twist soliton at $z=30$. As shown in the final states on the right, the in-plane current drives both longitudinal and transverse motion of the tube, resulting in final positions of $(x,y)=(50,28)$ for $\chi = -1$ and $(50,23)$ for $\chi = 1$ under periodic boundary conditions~\cite{note_circulation}. This behavior is consistent with conventional 2D skyrmion dynamics referred to as the skyrmion Hall motion~\cite{Jiang2017}. Besides these lateral motions, the in-plane current also drives out-of-plane motion of the twist solitons, as indicated by the red planes marking their final positions at $z = 37$ for $\chi = -1$ and $z = 24$ for $\chi = 1$. This suggests that the twist soliton chirality determines the direction of the out-of-plane motion.

To elucidate the mechanism underlying this motion, we examine its dependence on the current $j_{\rm e}$. The left panels of Figs.~\ref{DCjey}(b) and \ref{DCjey}(d) present the $j_{\rm e}$ dependences of the in-plane velocities $v_x$ and $v_y$ averaged over $\tau = 20\,000 - 50\,000$ for $\chi=-1$ and $1$, respectively. The circle markers obtained from the LLG simulations show a linear trend in both cases, with no discernible dependence on $\chi$. In contrast, the right panels show that the twist soliton velocity $v_z$ changes sign with $\chi$. Moreover, $v_z$ scales quadratically with $j_{\rm e}$, as shown by the purple solid line. These observations imply that the in-plane current induces an out-of-plane twist motion with $v_z \propto \chi j_{\rm e}^2$

The steady-state dynamics of the twisted tube can be analyzed using a collective-coordinate approach based on the Thiele equation (see End Matter for details). Within this description, we find that the dissipative tensor $\mathcal{D}_{ij}$ in Eq.~\eqref{eq:D} plays a key role in the out-of-plane motion. Specifically, the soliton velocity $v_z$ is given by
\begin{align}
  v_z = \left( \frac{\beta}{\alpha} - 1 \right) \frac{\mathcal{D}_{YZ}}{\mathcal{D}_{ZZ}} \frac{p j_{\rm e}}{2}.
  \label{eq:nonlinearvz}
\end{align}
The velocities obtained from Eq.~\eqref{eq:thiele} and its simplified form in Eq.~\eqref{eq:nonlinearvz} are overlaid in Figs.~\ref{DCjey}(b) and \ref{DCjey}(d). The Thiele-based results for $v_x$ and $v_y$ are in quantitative agreement with those directly obtained from the LLG simulations. In contrast, the analytical lines for $v_z$ deviate from the numerical results, although they capture the quadratic trend, thereby supporting the validity of Eq.~\eqref{eq:nonlinearvz}. This discrepancy may be attributed to the limitations of the rigid-body approximation in the Thiele-based analysis, as well as lattice-discretization effects in evaluating Eqs.~\eqref{eq:G} and \eqref{eq:D}.

We can show that $\mathcal{D}_{YZ}/\mathcal{D}_{ZZ}$ in Eq.~\eqref{eq:nonlinearvz} scales as $\chi j_{\rm e}$ by constructing an ansatz for the current-driven spin configuration as follows. Since $\mathcal{D}_{ZZ}$ is positive definite, we focus on $\mathcal{D}_{YZ}$. A classical spin field can generally be parametrized as $\bold{S}(\rho, \varphi, z)=(\sin \Theta \cos \Phi, \sin \Theta \sin \Phi, \cos \Theta)$, where $(\rho, \varphi, z)$ denotes cylindrical coordinates and $\Theta$ and $\Phi$ are their functions. The spatial dependence of $\Theta$ primarily controls the radius of the skyrmion tube. For simplicity, we assume an ansatz in which $\Theta$ depends only on $\rho$, namely $\Theta=f(\rho)$, with $f(0) = \pi$ and $f(\rho \to \infty) = 0$. $\Phi$ determines the skyrmion helicity and therefore requires a careful choice to describe the twist soliton motions. We adopt its ansatz as $\Phi = \varphi + \eta (z) + j_{\rm e} u(\varphi)$, where $\eta (z)$ is the helicity variation along the $z$ direction and satisfies $\int_0^{L_z} \frac{d\eta(z)}{dz}dz = 2\pi \chi$, with $L_z$ being the system size along the $z$ direction. Here, $u(\varphi)$ describes a first-order correction induced by the current and is taken to be a periodic function satisfying $u(0)=u(2\pi)$. Such current-induced helicity modulation was discussed in Ref.~\cite{Masell2020}.

For this ansatz, $\mathcal{D}_{YZ}$ is calculated up to first order in $j_{\rm e}$ as
\begin{align}
  \mathcal{D}_{YZ} &= 2\chi \pi j_{\rm e} \int\nolimits_0^{2\pi} d\varphi\,\cos\varphi \frac{du}{d\varphi} \int\nolimits_0^{\infty} d\rho\,\sin^2f \nonumber \\
  &= 2\chi \pi^2 j_{\rm e}c_{j_{\rm e}} \int\nolimits_0^{\infty} d\rho \sin^2f.
  \label{eq:DYZ}
\end{align}
The second line is obtained by expanding $u(\varphi)$ in Fourier series and denoting the coefficient of the first harmonic by $c_{j_{\rm e}}$. Plugging Eq.~\eqref{eq:DYZ} in Eq.~\eqref{eq:nonlinearvz} and considering $\beta/\alpha=1/2$, we find that the direction of the twist soliton motion arising from an in-plane current is determined by
\begin{align}
  \mathrm{sign}(v_z) = -\mathrm{sign}(\chi c_{j_{\rm e}}).
  \label{eq:signvz}
\end{align}

\begin{figure}[t!]
  \centering
  \includegraphics[width=\hsize]{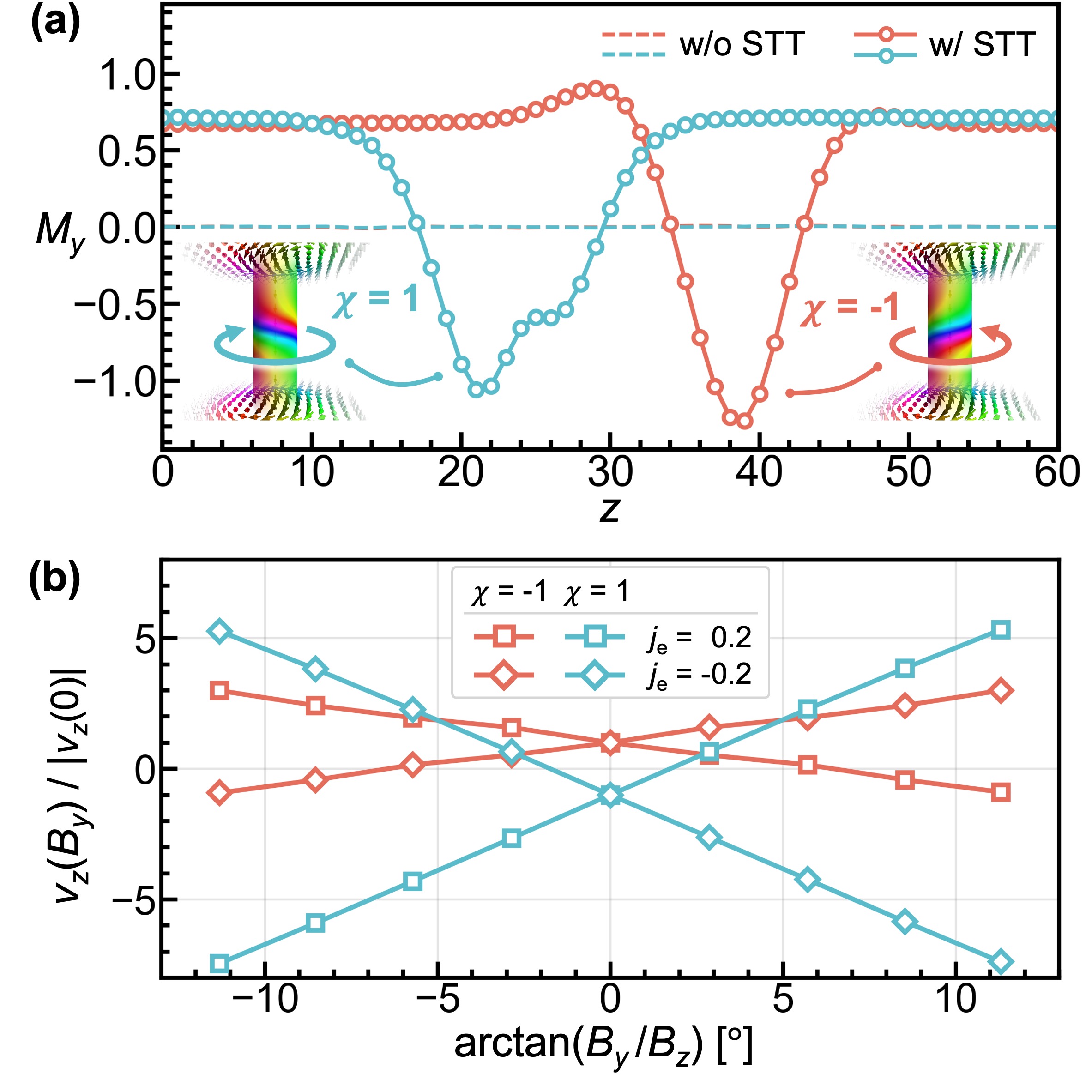}
  \caption{(a) Chirality dependence of the STT-induced magnetization $M_y$ at each $z$ under $j_{\rm e} = 0.2$. The red and blue data correspond to $\chi = -1$ and $\chi = 1$, respectively. The dashed and solid lines correspond to the cases without and with STT, respectively; the latter is taken at $\tau = 50\,000$. (b) Twist soliton velocity normalized by its value at $B_y = 0$ as a function of the tilt angle of the external magnetic field. We use $B_z = 0.02$ in both (a) and (b).}
  \label{my_chi2}
\end{figure}

To verify Eq.~\eqref{eq:signvz}, we evaluate the deformation of the skyrmion tube induced by the STT-induced sinusoidal mode by calculating the net magnetization along the $y$ direction. Assuming $\Phi = \varphi + \eta(z) + j_{\rm e}c_{j_{\rm e}} \sin \varphi$, the $y$ component of the spin configuration up to the first order in $j_{\rm e}$ is given by $S^y = (\sin(\varphi + \eta) + j_{\rm e}c_{j_{\rm e}}\cos (\varphi + \eta)\sin\varphi)\sin f$. Integrating $S^y$ over each $xy$ plane, the current-induced magnetization is obtained as $M_y(z) = -j_{\rm e}\pi c_{j_{\rm e}} \sin \eta(z) I$, where the integral $I = \int_0^{\infty} \rho \sin f \, d\rho$ is positive. We then introduce the difference in $M_y$ between the twisted position with $\eta = \pi/2$ and the intact region with $\eta = 3\pi/2$ [see also Fig.~\ref{schematic}(a)]:
\begin{align}
  \Delta M_y = M_{y}|_{\eta=\frac{\pi}{2}} - M_{y}|_{\eta=\frac{3\pi}{2}} = -2j_{\rm e}\pi c_{j_{\rm e}} I.
  \label{eq:DelMy}
\end{align}
Using this relation, Eq.~\eqref{eq:signvz} can be rewritten as
\begin{align}
  \mathrm{sign}(v_z) = \mathrm{sign}(\chi \Delta M_y j_{\rm e}).
  \label{eq:signvz_re}
\end{align}
Equation~\eqref{eq:signvz_re} shows that the direction of the out-of-plane motion is governed by the product of the twist chirality $\chi$, the STT-induced $\Delta M_y$, and current $j_{\rm e}$. Figure~\ref{my_chi2}(a) plots $M_y$ as a function of $z$ for the snapshots presented in Figs.~\ref{DCjey}(a) and \ref{DCjey}(c). In the initial state, $M_y$ vanishes due to the azimuthal symmetry of the skyrmion tube, as indicated by the dashed lines. After applying the current, nonzero and nonmonotonic $M_y$ profiles appear, with dips corresponding to the twist-soliton positions. Since $\Delta M_y$ is negative for both chiralities, Eq.~\eqref{eq:signvz_re} predicts motion in opposite $z$ directions for $\chi=-1$ and $\chi=1$, in agreement with the simulations.

The sign relation between the twist soliton motion and the STT-induced magnetization described in Eq.~\eqref{eq:signvz_re} suggests potential control of the motion via an in-plane magnetic field $B_y$ in the $y$ direction. Following the derivation of Eq.~\eqref{eq:DYZ}, the linear-response contribution of $B_y$ to $\mathcal{D}_{YZ}$ modifies Eq.~\eqref{eq:DYZ} as $\mathcal{D}_{YZ} \propto \chi (j_{\rm e}c_{j_{\rm e}} + B_y c_{B_y})$, where $c_{B_y}$ denotes the first-harmonic Fourier coefficient induced by $B_y$. Then, the sign relation in Eq.~\eqref{eq:signvz_re} is generalized with $\Delta M_y \propto c_{j_{\rm e}}j_{\rm e}+c_{B_y}B_y$. These results indicate that an in-plane magnetic field enables linear control of the soliton velocity and induces nonreciprocity through its coupling to $j_{\rm e}$.

Figure~\ref{my_chi2}(b) presents the twist soliton velocity normalized by its absolute value at $B_y = 0$ for each combination of chirality and current direction. The horizontal axis represents the tilt angle of the magnetic field measured from the out-of-plane direction. Since $B_y \ll B_z=0.02$, this angle is approximately $B_y/B_z$. The linear behavior of all curves and the nonreciprocity under sign reversal of $j_{\rm e}$ are consistent with the analytical prediction discussed above. We also confirmed that the analytical sign relation in Eq.~\eqref{eq:signvz_re} holds except for the data where the velocity becomes nearly zero (not shown). Importantly, we find that a small in-plane component, $B_y/B_z \sim \pm \ang{10}$, enhances the twist soliton velocity by several times and enables control of its direction of motion. The chirality dependence of this enhancement likely reflects the chirality-dependent susceptibility of the tube to structural deformation in the present spin model including the DMI.

Finally, we discuss an electrical scheme for detecting twist soliton motion via the emergent electric field (EEF). The EEF, also known as the spin motive force, is an effective electric field generated by the dynamics of spin textures, such as domain walls, spin helices, and skyrmions~\cite{Yang2009,Schulz2012,Yamane2019,Yokouchi2020}. Its $q$ component $(q = x,y,z)$ integrated over the space is defined as~\cite{Duine2008,Tserkovnyak2008,Shibata2011}
\begin{align}
  E^{\rm em}_q = \int \bold{S} \cdot \left( \frac{\partial \bold{S}}{\partial \tau} \times \frac{\partial \bold{S}}{\partial q}\right) \, d\bold{r} + \beta \int \left(\frac{\partial \bold{S}}{\partial \tau} \cdot \frac{\partial \bold{S}}{\partial q}\right) \, d\bold{r}.
  \label{eq:EEF_def}
\end{align}
Within the rigid-body approximation, a spin texture moving with a constant velocity $\bold{v}$ satisfies $\frac{\partial \bold{S}}{\partial \tau} = -(\bold{v} \cdot \bold{\nabla})\bold{S}$. Using Eqs.~\eqref{eq:G} and \eqref{eq:D}, Eq.~\eqref{eq:EEF_def} is written as $E^{\rm em}_q = \sum_{\mu = x,y,z} v_{\mu}(\mathcal{G}_{\mu q} - \beta \mathcal{D}_{\mu q})$, where the subscripts in $\mathcal{G}_{\mu q}$ and $\mathcal{D}_{\mu q}$ should be understood as uppercase collective-coordinate indices. We consider the $z$ component of the EEF generated by twist soliton motion driven by an in-plane current along the $y$ direction. In the continuum limit, the gyrotropic contribution vanishes because the skyrmion numbers on both the $xz$ and $yz$ cross sections are zero. The remaining dissipative contribution is therefore given by
\begin{align}
  E^{\rm em}_z = -\beta (v_x\mathcal{D}_{XZ} + v_y\mathcal{D}_{YZ} + v_z\mathcal{D}_{ZZ}).
  \label{eq:EEF_D}
\end{align}
Since $\mathcal{D}_{ZZ} \gg \mathcal{D}_{YZ} \sim \mathcal{D}_{XZ}$ (see End Matter) and $v_y \gg v_x \sim v_z$ as shown in Figs.~\ref{DCjey}(b) and \ref{DCjey}(d), the second and third terms in Eq.~\eqref{eq:EEF_D} dominate the EEF. Furthermore, since $\mathcal{D}_{YZ} \propto \chi j_{\rm e}, v_z \propto \chi j_{\rm e}^2,$ and $v_y \propto j_{\rm e}$, while $\mathcal{D}_{ZZ}$ is independent of both $\chi$ and $j_{\rm e}$, the EEF scales as $\chi j_{\rm e}^2$. 

\begin{figure}[t!]
  \centering
  \includegraphics[width=\hsize]{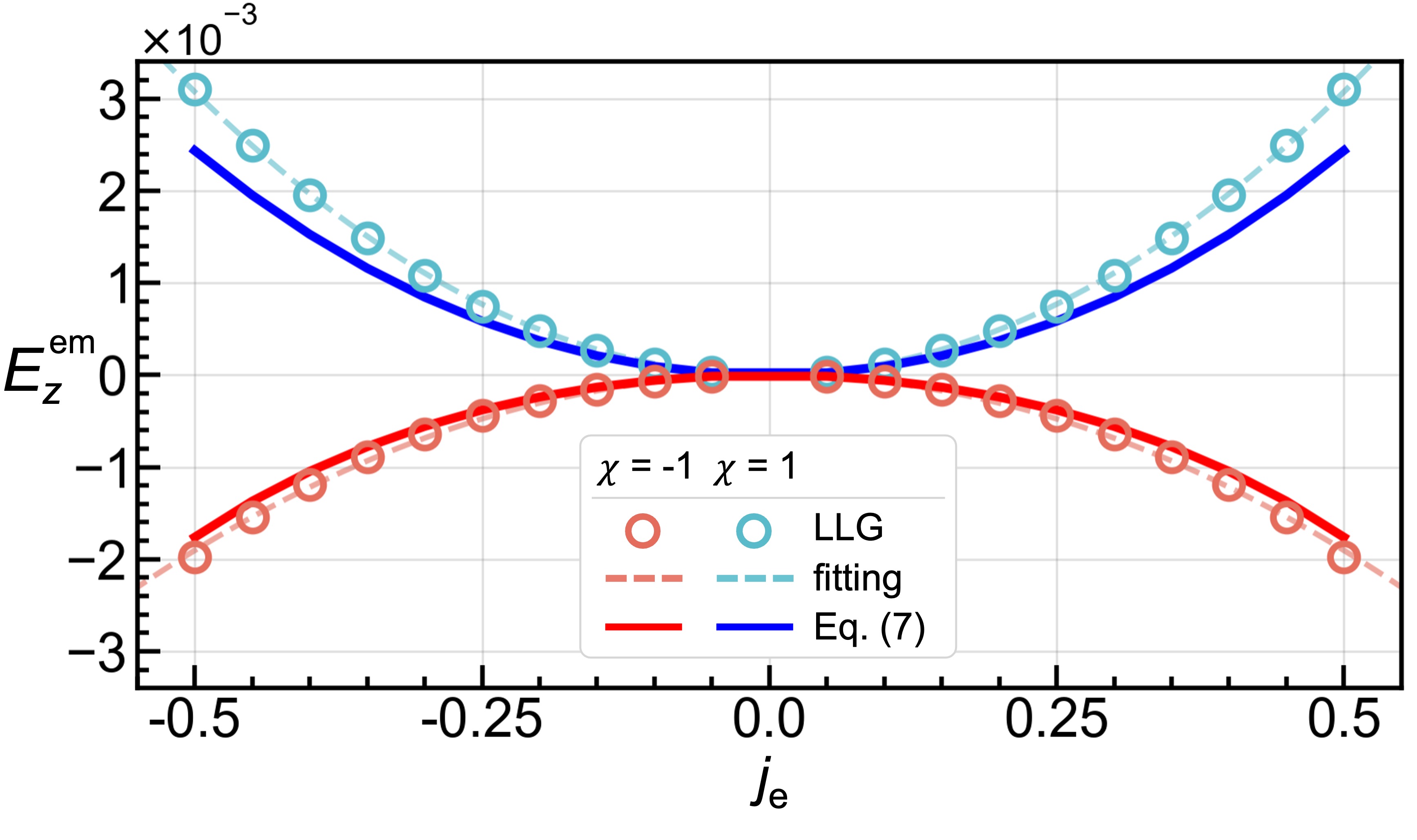}
  \caption{$\chi$ and $j_{\rm e}$ dependence of the nonadiabatic contribution to the EEF with $\beta = 0.02$ and $B_z = 0.02$. The circles represent the values obtained by directly evaluating the second term in Eq.~\eqref{eq:EEF_def} from snapshots of the LLG dynamics, while the dashed lines show fits $\propto$ $j_{\rm e}^2$. The solid lines are the Thiele-based predictions obtained from Eq.~\eqref{eq:EEF_D}.}
  \label{Eem_beta}
\end{figure}

Figure~\ref{Eem_beta} shows the $z$ component of the second term of Eq.~\eqref{eq:EEF_def} as a function of $j_{\rm e}$. The circles represent the values obtained by directly evaluating the time derivative in Eq.~\eqref{eq:EEF_def} using the LLG equation in Eq.~\eqref{eq:LLG}, and are fitted by the dashed lines proportional to $j_{\rm e}^2$. The twist-chirality dependence of the sign of the EEF and the good agreement with the quadratic fits are consistent with the discussion above. The solid lines obtained from Eq.~\eqref{eq:EEF_D}, using $\mathcal{D}$ and $\bold{v}$ calculated from Eq.~\eqref{eq:thiele}, reproduce the same trend as the numerical results. This chirality-dependent EEF suggests that twist solitons, including their chirality, can be electrically detected through out-of-plane Hall-voltage measurements.

To summarize, we have identified a twist soliton formed on a skyrmion tube and clarified its formation mechanism, current-driven dynamics, and potential detection scheme. We showed that twisted skyrmion tubes are created probabilistically during relaxation dynamics that mimics a thermal quench. We further found that an in-plane electric current induces chirality-dependent out-of-plane motion of the twist solitons via STT-induced magnetization, with a velocity that can be strongly amplified by a slight tilt of the applied magnetic field. Finally, we proposed an electric detection scheme based on out-of-plane Hall voltage measurements.

While this work focused on a singly twisted skyrmion tube, crystalline states of multiply twisted tubes warrant further investigation, as a measurable Hall-voltage may require the collective contributions of many twist solitons. Developing more direct and deterministic creation protocols, such as twist injection from a sample edge using tailored electromagnetic fields, is another important challenge~\cite{Fujita2017, Knapman2024, Shimizu2025}. The effect of twist solitons on magnetic excitations also merits further investigation. Moreover, the intimate connection between twisted tube and knot topology suggests that these research directions may shed light on the physics of hopfions~\cite{azhar2024,Shaju2026}. More broadly, our results may provide insight into the formation and dynamics of twisted magnetic flux tubes in astrophysical environments, which are believed to play an important role in solar-flare activity~\cite{Priest2002}. Our findings establish the twist degree of freedom of skyrmion tubes as a promising platform for 3D topological spin physics and device functionalities.

%%%%%%%%%%%%%%%%%%%%%%%%%%%%%%%%%%%%%%%%%%%%%%%%%%%%%%%%%%%%%%%%%%%%%%%%%%%%%%%%%%%%%%%%%%%%%%%%%%%%%%%%%%%%%%%%%%%%%%%%%%%%%%%%%%%%

% \section*{\label{acknowledge}Acknowledgments}
This work was supported by the JSPS KAKENHI (No.~22K13998, No.~23K25816, No.~JP24K22870, No.~JP25H01247, and No.~26H00634) and JST PRESTO (No.~JPMJPR2595). S. K. was supported by the Program for Leading Graduate Schools (MERIT-WINGS) and JST SPRING, Grant Number JPMJSP2108. The computation in this work has been done using the facilities of the Supercomputer Center, the Institute for Solid State Physics, The University of Tokyo.

%%%%%%%%%%%%%%%%%%%%%%%%%%%%%%%%%%%%%%%%%%%%%%%%%%%%%%%%%%%%%%%%%%%%%%%%%%%%%%%%%%%%%%%%%%%%%%%%%%%%%%%%%%%%%%%%%%%%%%%%%%%%%%%%%%%%

\bibliography{bibliography}

\appendix

\newpage
\onecolumngrid
\vspace{\columnsep}
\begin{center}
\rule[3pt]{0.4\textwidth}{0.4pt}
\textbf{\large{ \ End Matter \ }}
\rule[3pt]{0.4\textwidth}{0.4pt}
\end{center}
\vspace{\columnsep}
\twocolumngrid

{\it Model and Method}
---We introduce the model Hamiltonian and numerical methods used in this study. We consider a 3D chiral magnet with competing spin interactions on a simple cubic lattice. The Hamiltonian is given by
\begin{align}
  \mathcal{H} = -\sum_{\alpha = 1}^4 \sum_{\langle i,j \rangle_\alpha} &J_{\alpha}~\bold{S}_i \cdot \bold{S}_j + \sum_{\langle i,j \rangle_1} \bold{D}_{i,j} \cdot (\bold{S}_i \times \bold{S}_j) \nonumber \\
  & - \bold{B} \cdot \sum_{i}\bold{S}_{i} - K \sum_i (S^z_i)^2,
\label{eq:model}
\end{align}
where $\bold{S}_i = (S_i^x,S_i^y,S_i^z)$ represents the classical spin with $|\bold{S}_i|=1$ at site $i$. The first term represents the exchange interactions up to the fourth-neighbor pairs; the summation with respect to $\langle i,j \rangle_{\alpha}$ runs over $\alpha$th-neighbor pairs. We take $(J_1,J_2,J_3,J_4) = (1,-0.164,0,-0.082)$, similar to the parameters used in previous studies on hopfions~\cite{Bogolubsky1988,Liu2020,Rybakov2022}. The magnetic frustration is introduced to circumvent the limitation imposed by Derrick's theorem~\cite{Derrick1964}. The second term denotes the DMI between nearest-neighbor pairs. The DM vector is taken along the bond direction $\bold{D}_{i,i+\mu} = D\hat{\bm \mu}$, where $\hat{\bm \mu}$ is the unit vector along the $\mu = x, y, z$ direction. We set $D = 0.01$, which lifts the helicity degeneracy and stabilizes the $\eta = 3\pi/2$ state as the lowest-energy skyrmion state. The third term describes the Zeeman coupling to the magnetic field $\bold{B}=(0,B_y,B_z)$, with $B_y = 0$ unless otherwise noted. The fourth term denotes the single-ion anisotropy; positive (negative) $K$ prefers out-of-plane (in-plane) spin direction. We set the lattice constant to unity and impose periodic boundary conditions.

To investigate real-space and real-time spin dynamics, we solve the LLG equation given in dimensionless form,
\begin{align}
  \frac{d \bold{S}_i}{d\tau} &= \frac{1}{1 + \alpha^2} (\bold{S}_i \times \bold{H}_i^{\rm eff} + \alpha \bold{S}_i \times (\bold{S}_i \times \bold{H}_i^{\rm eff}) + \bold{T}_i),
  \label{eq:LLG}
\end{align}
where $\tau$ and $\alpha$ represent the dimensionless time and the Gilbert damping constant, respectively. The effective field is defined as $\bold{H}_i^{\rm eff} = \frac{\partial \mathcal{H}}{\partial \bold{S}_i}$. The STT induced by an electric current is given by~\cite{Zhang2004}
\begin{align}
  \bold{T}_i = (\beta& - \alpha)\bold{S}_i \times (\bold{j}_{\rm e} \cdot \bm{\nabla}) \bold{S}_i \nonumber \\
  &+ \alpha \beta \bold{S}_i \times [\bold{S}_i \times (\bold{j}_{\rm e} \cdot \bm{\nabla}) \bold{S}_i] - (\bold{j}_{\rm e} \cdot \bm{\nabla}) \bold{S}_i,
\label{eq:STT}
\end{align}
where $\bold{j}_{\rm e} = p j_{\rm e}\hat{\bm{\nu}}/2$, $p$ is the spin polarization of the current, $j_{\rm e}$ is the current amplitude, and $\hat{\bm{\nu}}$ is a unit vector describing the current direction. $\beta$ is the nonadiabatic coefficient. On the discrete lattice, the spatial derivative $\bm{\nabla} = (\partial_x, \partial_y, \partial_z)$ is computed as $\partial_{\nu} \bold{S}_i = (\bold{S}_{i + \nu} - \bold{S}_{i - \nu})/2$. In the continuum limit, $\bm{\nabla}$ becomes the ordinary nabla operator, and the second term in Eq.~\eqref{eq:STT} reduces to $-\alpha \beta (\bold{j}_{\rm e} \cdot \bm{\nabla}) \bold{S}_i$ because $|\bold{S}_i| = 1$ and $\bold{S}_{i} \cdot (\bold{j}_{\rm e} \cdot \bm{\nabla}) \bold{S}_i = 0$. We numerically integrate Eq.~\eqref{eq:LLG} using the fourth-order Runge-Kutta method with time step $\Delta \tau = 0.1$. Throughout this work, we set $p = 0.2$, $\alpha = 0.04$, and $\beta = 0.02$, which are typical values for ferromagnetic metals~\cite{Zhang2004, Oogane2006}.

{\it Formation rate of skyrmion tubes}
---We present statistics of the topological structures created in 500 independent quench dynamics simulations starting from random initial spin configurations. We find that twisted tubes with $(N_{\rm sk}, \chi) = (-1, -1)$, which are topologically equivalent to that in Fig.~\ref{quench}, are generated 13 times. Here, $N_{\rm sk}$ denotes the skyrmion number on a horizontal cross section of the texture. Among these events, one realization exhibits two tubes within the system. In addition, with lower probabilities, tubes with $(N_{\rm sk}, \chi) = (-1, 0)$ and $(N_{\rm sk}, \chi) = (1, 1)$ are observed four and two times, respectively. This preference appears to depend on the choice of the DM vectors in Eq.~\eqref{eq:model}. These results suggest that twists on skyrmion tubes can be stabilized in skyrmion-hosting cubic chiral magnets such as MnSi and FeGe~\cite{Muhlbauer2009, Yu2011}.
A more detailed statistical analysis of twist-tube formation during anneal, aimed at investigating connections with KZM physics, is left for future work.

{\it Collective-coordinate approach}
---We introduce the Thiele equation, which describes the steady-state motion of magnetic textures, and show that it reduces to Eq.~\eqref{eq:nonlinearvz} in the weak current regime. For simplicity, we assume that the dynamics of the twisted tube is governed solely by rigid translational motion along the $x, y,$ and $z$ directions. The spin configuration at time $\tau$ is then written as $\bold{S}(\bold{r}, \tau) = \bold{S}(\bold{r} - \bold{v} \tau, 0)$. Here, $\bold{v}$ denotes the steady-state velocity of the twisted tube, which is given by the time derivative of the collective coordinate $\bold{q} = (X(\tau), Y(\tau), Z(\tau))$ specifying the center position of the structure, i.e., $\bold{v}=\dot{\bold{q}}$. The equation of motion for the twisted tube is then given by the Thiele equation~\cite{Masell2021}
\begin{align}
  \dot{\bold{q}} = (\mathcal{G} + \alpha \mathcal{D})^{-1} (\mathcal{G} + \beta \mathcal{D}) \bold{j}_{\rm e},
  \label{eq:thiele}
\end{align}
where $\mathcal{G}$ and $\mathcal{D}$ are the gyrotropic and dissipative tensors, respectively, with
\begin{align}
  \mathcal{G}_{i j} &= -\int \bold{S} \cdot\left(\frac{\partial \bold{S}}{\partial q_{i}} \times \frac{\partial \bold{S}}{\partial q_{j}}\right) d \bold{r},\label{eq:G}\\
  \mathcal{D}_{i j} &= \int \left(\frac{\partial \bold{S}}{\partial q_{i}} \cdot \frac{\partial \bold{S}}{\partial q_{j}}\right) d \bold{r}.
  \label{eq:D}
\end{align}
Since Eq.~\eqref{eq:G} represents the integral of the skyrmion number density over the $ij$ plane, only $\mathcal{G}_{XY} = -\mathcal{G}_{YX}$ are finite in the present case. For Eq.~\eqref{eq:D}, we assume that the in-plane current lowers the azimuthal symmetry of the skyrmion tube, thereby allowing the off-diagonal components of $\mathcal{D}$ to take nonzero values, although they remain much smaller than the diagonal components. Indeed, for the texture shown in Fig.~\ref{DCjey}(a), we numerically obtain $\mathcal{G}_{XY} = 4\pi \times 61 \simeq 767, \mathcal{D}_{XX} \simeq 732, \mathcal{D}_{YY} \simeq 732, \mathcal{D}_{ZZ} \simeq 77, \mathcal{D}_{XY} \simeq -0.02, \mathcal{D}_{XZ} \simeq -0.41,$ and $\mathcal{D}_{YZ} \simeq -1.86$. We have also confirmed that the $\chi = 1$ twisted tube in Fig.~\ref{DCjey}(c) exhibits comparable values to them. Using these tensor components, the soliton velocity $v_z$, the third component of Eq.~\eqref{eq:thiele}, is expressed as
\begin{widetext}
\begin{align}
v_z &= \left( \frac{\beta}{\alpha} - 1 \right)
\frac{\mathcal{G}_{XY}
\left[
\mathcal{D}_{YZ} \mathcal{G}_{XY}
+ \alpha (\mathcal{D}_{XZ} \mathcal{D}_{YY}
- \mathcal{D}_{XY} \mathcal{D}_{YZ})
\right]}
{\mathcal{D}_{ZZ} \mathcal{G}_{XY}^{2} - \alpha^{2}\left(
\mathcal{D}_{YY} \mathcal{D}_{XZ}^{2}
- 2 \mathcal{D}_{XY} \mathcal{D}_{XZ} \mathcal{D}_{YZ}
+ \mathcal{D}_{XX} \mathcal{D}_{YZ}^{2}
+ \mathcal{D}_{ZZ} \mathcal{D}_{XY}^{2}
- \mathcal{D}_{XX} \mathcal{D}_{YY} \mathcal{D}_{ZZ}
\right)} \frac{p j_{\rm e}}{2}.
\label{eq:vz_thiele}
\end{align}
\end{widetext}
Given the small off-diagonal components of $\mathcal{D}$ and small $\alpha$, Eq.~\eqref{eq:vz_thiele} reduces to Eq.~\eqref{eq:nonlinearvz} in the main text.

\end{document}